\documentclass[pre,twocolumn,a4paper,superscriptaddress]{revtex4}
\usepackage{epsfig}
\begin{document}

\title{Multifractal analysis of DNA walks and trails }
\author{Alexandre Rosas}
\affiliation{Instituto de F\'{\i}sica de S\~ao Carlos, Universidade de S\~ao
Paulo, Caixa Postal 369, 13560-970 S\~ao Carlos SP, Brazil.}
\author{Edvaldo Nogueira Jr.}
\affiliation{Instituto de F\'{\i}sica de S\~ao Carlos, Universidade de S\~ao
Paulo, Caixa Postal 369, 13560-970 S\~ao Carlos SP, Brazil.}
\affiliation{Instituto de F\'{\i}sica, Universidade Federal da Bahia, Campus da
Federa\c{c}\~ao, 40210-340, Salvador, BA, Brazil.}
\author{Jos\'e F. Fontanari}
\affiliation{Instituto de F\'{\i}sica de S\~ao Carlos, Universidade de S\~ao
Paulo, Caixa Postal 369, 13560-970 S\~ao Carlos SP, Brazil.}

\begin{abstract}

The characterization of  the long-range order and fractal properties of
DNA sequences 
has proved a difficult though rewarding task due mainly to 
the mosaic character of DNA  consisting of many
interwoven patches of various lengths with different nucleotide constitutions.
We apply here a recently proposed generalization of the detrended fluctuation analysis 
method to show that the DNA walk construction, in which 
the DNA sequence is viewed as a time series, exhibits a monofractal structure 
regardless of the existence of  local trends in the series.
In addition,
we point out that the monofractal structure of the DNA walks carries over to
an apparently alternative graphical construction given by the projection of the DNA walk 
into the $d$ spatial coordinates, termed DNA trails. In particular, we 
calculate  the fractal dimension $D_t$ of the DNA trails using
a well-known result of fractal theory linking $D_t$ to the Hurst exponent  
$H$ of the corresponding DNA walk. Comparison with  estimates obtained
by the standard box-counting method allows the evaluation of both  
finite-length and local trends effects.
 
\end{abstract}


\maketitle

\section{Introduction}\label{sec:1}

The search for patterns in nature and their interpretation in terms of 
general principles is one of the main purposes of
science. The explosive accumulation of DNA sequence data in
the last two decades has provided a rich source of raw material 
that hides valuable hints about the evolutionary mechanisms
of the genome organization. Unveiling the patterns in those
sequences has become an exciting challenge to the present generation of
statistical physicists and information scientists.
In that vein, a truly remarkable result was the finding that intron-containing DNA 
coding regions exhibit 
long-range power-law correlations 
extending across more than $10^4$ nucleotides,
whereas intronless coding regions display only short-range correlations 
\cite{Peng92,Peng94}.

Characterizing the long-range correlations in DNA sequences is a highly nontrivial 
task because of the mosaic structure of DNA consisting of patches with different nucleotide
composition \cite{Karlin93}. In fact, in order to use the standard techniques 
(e.g., power spectrum and R/S fluctuation analysis) to study
the correlations of DNA sequences, it is necessary first
to eliminate the local biases in nucleotide composition (trends), thus avoiding  
spurious effects due to the mosaic character  of the sequence.
In the context of time series or records there are two main techniques to eliminate 
these  trends, namely, the 
detrended fluctuation analysis (DFA) \cite{Peng94} and the wavelet transform (WT)
\cite{Chui92}. The former is a physically
appealing {\it ad hoc} technique of easy implementation and the latter is a mathematically
well-established transform used to study the regularity of arbitrary functions via 
the systematic elimination of local polynomial behavior. Both techniques can be readily
applied to the analysis of ordered linear sequences, such as DNA, by considering the 
reading direction as the time axis (see e.g. \cite{Peng94,Galvan96} for the DFA
and \cite{Arneodo95,Tsonis96} for the WT analyses of DNA sequences).

A common strategy to represent graphically a given DNA
sequence consists of transforming it into a $d$-dimensional random walk,
so-called DNA walk, by associating a space direction to each nucleotide (A,C,T,G) or class of
nucleotides (e.g. purine and pyrimidine) and the time direction to the reading
direction of the sequence \cite{Peng92}. Here, A, C, T and G are the bases adenine, cytosine,
thymine, and guanine, respectively. In particular, a 
very popular representation
is the one-dimensional walk in which a purine (A or G) at position $i$ is associated to
one step down ($y_i = -1$) while a pyrimidine (T or C) is associated to one
step up ($y_i = 1$). We note that the purine-pyrimidine classification is
related to the hydrophobic-hydrophilic characteristics of the encoded amino acids.
Of course, there are many alternative DNA walks and the most complete one that considers
the four bases equivalently in base space is the  three-dimensional DNA walk \cite{Luo98},
in the sense that  its projections on appropriate  axes or planes recover all possible
one and two-dimensional walks.

\begin{figure}
\centerline{\epsfig{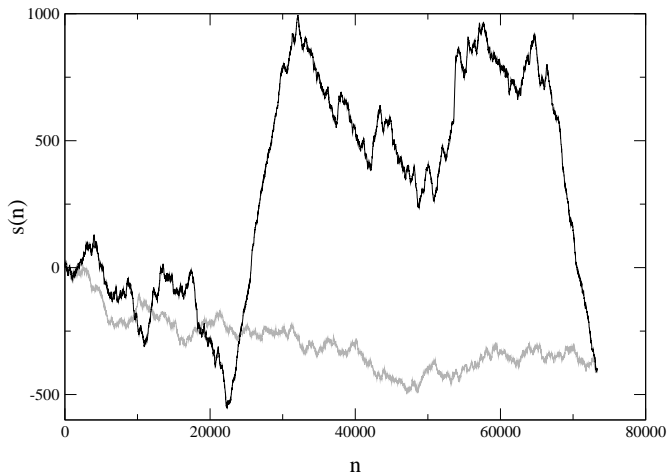}}
\par
\caption{Purine-pyrimidine random walk plot of the  human beta-globin region
(black line) and a shuffled sequence of the same nucleotide
composition (grey line).}
\label{fig:1}
\end{figure}

For the sake of concreteness, in this section we will focus on the fluctuations 
of the cumulative variable  of the  purine-pyrimidine random walk  only,
defined as 
\begin{equation}\label{profile}
s (n) = \sum_{i=1}^n y_i ~~~~n=1, \ldots, N,
\end{equation}
where $N$ is the length of the sequence.
This variable is displayed in Fig. \ref{fig:1} for the intron-rich human 
beta-globin region (GenBank name HUMHBB) for which $N= 73326$ as well as for a 
shuffled sequence with same length and nucleotide composition. The statistical quantity
that characterizes the fluctuations of $s(n)$ is the root mean square fluctuation
$F_2(l)$ defined as
\begin{equation}\label{def_var}
F_2 \left ( l \right ) = \left \{ \overline{\left [ \Delta s \left ( l \right ) \right ]^2}
-  \left [~\overline{\Delta s \left ( l \right ) } ~\right ]^2 \right \}^{1/2}
\end{equation}
where $\Delta s(l) = s(n+l) - s(n)$ and the bars indicate an average over all
positions $n$ in the record. We expect $F(l)$ to increase with increasing $l$,
the length of the window considered. Explicitly, in the
case of stationary series the fluctuations are described by a power law
\begin{equation}\label{power_law}
F_2 \left ( l \right ) \sim l^H 
\end{equation}
where the scaling exponent  (Hurst exponent)  $H$ equals $1/2$ in the case of 
random and short-ranged 
correlated sequences. Any value different from $1/2$ is evidence of the
existence of long-range correlations in the sequence \cite{Mandelbrot82,Voss89,Feder88}. 
In addition,
for self-affine records (e.g. DNA walks) there is a simple relation between
$H$ and the {\it local} fractal dimension of the record, $D_r = 2 - H$.
We note that the power law behavior implies that there is no characteristic length scale, 
i.e., very large fluctuations are likely due to the same mechanisms as smaller ones.

Complex records are unlikely to be fully characterized by a single scaling exponent as
evidenced by the case where the scaling behavior is different in distinct parts of
the series so that $H$ is actually dependent on the part of the 
record being analyzed.
A complete framework to describe the situation where a multitude
of scaling exponents is required is provided by the multifractal formalism
\cite{Feder88,Halsey86} and, in this sense, the multifractal spectrum may be viewed as
the ultimate tool to characterize a stationary time series. However 
the direct calculation of $F_2(l)$ using Eq. (\ref{def_var}) or the application of the 
standard multifractal formalism yield wrong results for nonstationary time series that
are affected by local trends. In this contribution we apply a generalization of the DFA
recently proposed by Kantelhardt {\it et al.} for the characterization of 
nonstationary time series to study the multifractal properties of DNA walks
\cite{Kantelhardt02}.  In agreement with a previous multifractality analysis
based on the wavelet transform modulus maxima (WTMM) method \cite{Arneodo95},
we find that the DNA walks exhibit a simple monofractal scaling behavior. The
multifractality analysis of one-dimensional DNA-walks is the subject of 
Sec. \ref{sec:2}.

\begin{figure}
\centerline{\epsfig{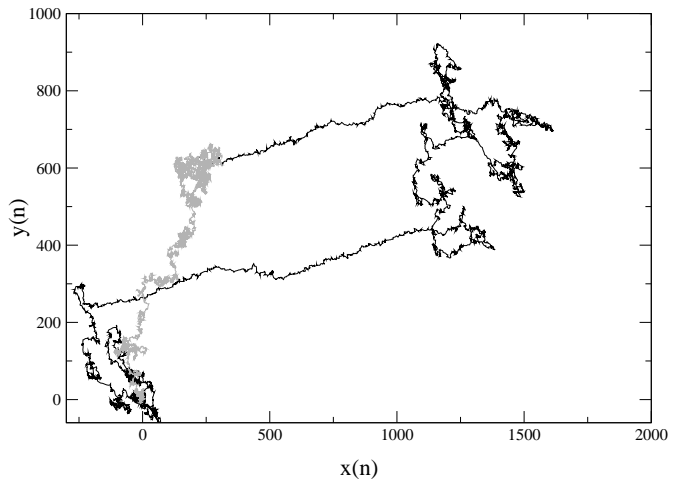}}
\par
\caption{Trail of the  human beta-globin region
(black line) and a shuffled sequence of the same nucleotide
composition (grey line) using the two-dimensional representation
T ($x_i = 1$), A ($x_i = -1$), G ($y_i = 1$) and C ($y_i = -1$).
\label{fig:2}}
\end{figure}

A seemingly alternative graphical representation of DNA sequences that
has received less attention and so less criticism than the aforementioned DNA walks 
is the so-called DNA pseudo random-walks 
\cite{Gates86,Berthelsen92,Berthelsen94,Glazier95,Oiwa02}.
They are simply the projection of the DNA walks into their $d$ spatial components, i.e.,
the {\it trails} of the DNA walks. Of course, drawing these trails makes sense
for $d \geq 2$ only, and Fig. \ref{fig:2} illustrates such a trail for a $d=2$
representation of the HUMHBB sequence
where right, left, up and down steps correspond
to the presence of nucleotides T, A, G and C, respectively. This 
representation preserves some basic symmetries of the DNA such as
complementarity, reflection, substitution and compatibility \cite{Berthelsen92}.

Despite the clear visual evidence of local trends  in the trails
exhibited  in Fig. \ref{fig:2}, it has been a common practice to apply standard
fractal and multifractal  methods (e.g., box counting \cite{Mandelbrot82} and 
sandbox \cite{Vicsek92}) to characterize their shapes  without 
much concern about the mosaic structure of the underlying DNA sequence.
Actually the efforts focused on the understanding of the disturbing  effects of the
finite length of the sequences on the estimate of the fractal dimension
$D_t$ as well as on the multifractal characterization of the trail. 
For instance, although it is well-known that in two-dimensions an infinite length  
true random  walk (i.e., a sequence of bases generated at random)
is space filling and so  $D_t=2$, the box-counting and sandbox methods 
yield estimates significantly lower than 2 even for walks as large as $10^5$
steps \cite{Berthelsen92}. 
Moreover, finite-length random walks exhibit an effective multifractal spectrum 
due mainly to crossover effects between the usually distinct bulk and surface properties of
the trail  \cite{Berthelsen94}. These caveats, however, have not prevented 
claims that
relatively short (typically $2 \times 10^4$ bp) mitochondrial DNA genomes have a definite
multifractal structure \cite{Oiwa02}. In Sec. \ref{sec:3} of this contribution
 we invoke a classical result of fractal theory to link the Hurst exponent 
of the DNA walk to the fractal
dimension of the corresponding trail in space (see, e.g., \cite{Mandelbrot82,Voss89}),
then arguing in favor of the monofractality of DNA trails too.

\section{DNA walks}\label{sec:2}

Consider a record, such as that given in Eq. (\ref{profile}), 
where the variables $y_i$
are not necessarily Ising variables but must form a compact support, i.e., $y_i =0$
for a very small fraction of the elements only. 
The generalized multifractal DFA (MF-DFA) involves the following steps
(we refer the reader to ref. \cite{Kantelhardt02} for a thorough presentation
of the method). First, divide the entire
sequence into $N_l = {\mathrm{int}}(N/l)$ nonoverlapping segments of length $l$.
Second, for each segment $\nu = 1, \ldots, N_l$ of length $l$ 
calculate the local trend by a least-square fit of the record in the segment, and
denote by $s_{\nu}(n)$ the fitting polynomial in segment $\nu$. Then evaluate
the variance
\begin{equation}\label{variancia}
F^2 \left ( l,\nu \right ) = \frac{1}{l} \sum_{n=1}^l \left \{ s \left [ \left
( \nu - 1 \right ) l + n \right ] - s_{\nu} (n) \right \}^2 
\end{equation}
for each segment. An important parameter is the order $m=0,1,2, \ldots$ of the polynomial
$s_{\nu}(n)$ used in the fitting procedure.  
The choices of $m$ correspond to different orders of DFA, denoted by DFA$m$,
which differ in their capability of eliminating trends in the series. For instance,
DFA$0$ can eliminate only constant trends in the series, DFA$1$ eliminates constant as
well as linear trends, 
and so on, so that a comparison of the results for different orders of DFA yields information
on the type of trend in the series. Third, determine the order $q$ fluctuation function 
averaging over all segments
\begin{equation}\label{F_q}
F_q \left ( l \right ) = \left \{ \frac{1}{N_l} \sum_{\nu=1}^{N_l} 
\left [ F^2 \left ( l,\nu \right ) \right ]^{q/2} \right \}^{1/q},
\end{equation}
where in general the index variable $q$ takes on any real value, except zero.
As the final step, analyze log-log plots of $F_q (l)$ versus $l$ for each value of $q$
in order to determine the scaling behavior of the fluctuation functions. For large values
of $l$, we expect these
functions to increase with increasing $l$ as a power law
\begin{equation}\label{pow_multi}
F_q \left ( l \right ) \sim l^{h(q)}.
\end{equation}
Clearly, for $q=2$ one has $h(2) = H$ by construction, and in the
case that the scaling behavior of $F^2 (l, \nu)$ is identical for all segments
$\nu$, i.e., the record is a monofractal, the exponent $h(q)$ is independent
of $q$, as expected. The definition (\ref{F_q}) is well-suited to detect
discrepancies in the scaling behavior of the large and small fluctuations. 
In particular, for positive $q$, $h(q)$ yields the scaling behavior of the segments with large
fluctuations  while for negative $q$, $h(q)$ yields
the scaling of the segments with small variances.
In addition, there is a simple relation between $h(q)$
and the scaling exponents $\tau (q)$ defined by the standard partition 
function-based multifractal formalism \cite{Kantelhardt02},
\begin{equation}\label{tau}
\tau (q) = q h(q) -1.
\end{equation}
The importance of this relation  is that it allows comparison of the  results
obtained in the MF-DFA scheme with those of the standard multifractal analysis in the case
of stationary series, and with those of the WTMM  in the case
of nonstationary series.

\begin{figure}
\centerline{\epsfig{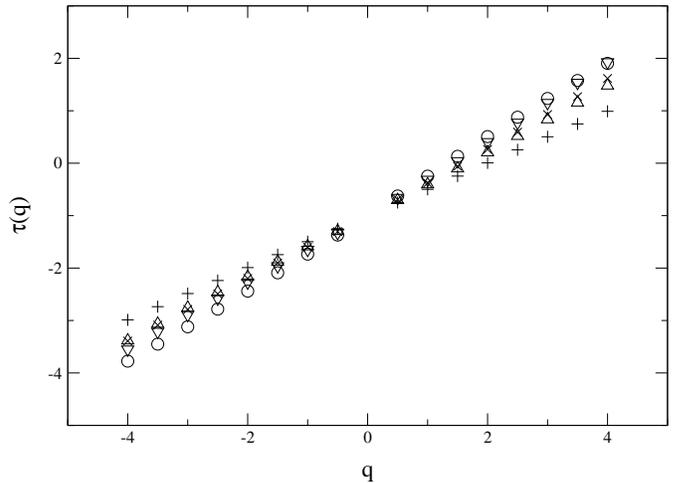}}
\par
\caption{Multifractal spectrum $\tau(q)$ vs $q$ for the DNA walk of the
intron-rich HUMHBB sequence depicted in Fig. \ref{fig:1}. The
data correspond to different orders of DFA, $m = 0 ~( \bigcirc ),
1~( \bigtriangledown ), 3~( \times )$  and $5~ ( \bigtriangleup )$.
The symbol $+$ corresponds to the data for the shuffled sequence.}
\label{fig:3}
\end{figure}

In Fig. \ref{fig:3} we show the multifractal spectrum $\tau(q)$ obtained by
the application of  different orders of the MF-DFA method to the intron-rich HUMHBB sequence. 
The results for orders larger than $5$ are indistinguishable in the scale of
the figure, so the data for $m=5 ~(\bigtriangleup )$ yields the trendless spectrum. 
Analysis of this figure gives valuable information on the nature of
the correlations of the DNA walk. In particular,
the linearity of the  trendless spectrum indicates that the record $s(n)$ depicted
in Fig. \ref{fig:1} is monofractal. Actually, this conclusion is not affected
by the presence of trends in the series, since  the $q$ dependence of
$\tau(q)$ is linear regardless of the value of $m$. However, the value of the scaling
exponent $H = h(q) ~\forall q$, given by the slope of the
straight lines that fit the data, is sensitive to local trends. For example,
we find $H = 0.72 \pm 0.02$ for $m=0$,  $H = 0.68 \pm 0.02$ for $m=1$,
and $H = 0.60 \pm 0.02$ for $m=5$. As expected, $H = 0.50 \pm 0.02$ for 
the shuffled sequence. All these numerical values are in perfect agreement with
the values obtained using the WTMM method \cite{Arneodo95}.
Since the trendless scaling exponent is definitively 
different from $1/2$ 
we can conclude for the existence of long-range correlations in this particular 
intron-containing sequence. The  long-range 
correlation induced by the local trends is reflected in the larger value
of the scaling exponent calculated with the MF-DFA$0$, which  overestimates 
the value of $H$ in about  $20\%$. It is interesting to note that the
trends influencing the large fluctuations ($q > 0$) are practically unaffected
by the application of DFA$1$.

\begin{figure}
\centerline{\epsfig{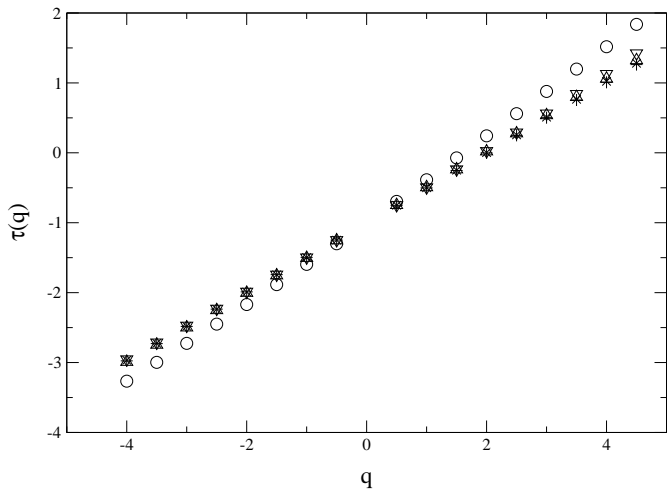}}
\par
\caption{Same as Fig. \ref{fig:3} but for the DNA walk of the
{\it E. coli} chromosomal sequence  ECO110K composed primarily
of coding regions.}
\label{fig:4}
\end{figure}

The results of the application of the MF-DFA method to the characterization of a
sequence composed predominantly of coding regions is summarized in Fig. \ref{fig:4}. 
The sequence considered is a portion of the {\it E. Coli} K12 genome
(GenBank name ECO110K)
for which $N=111401$.
We find $H = 0.60 \pm 0.02$ for $m=0$,  and $H = 0.51 \pm 0.02$ for $m \geq 1$
as well as for the shuffled sequence.
As before, the monofractal  structure of the record is confirmed
by the linearity of the $\tau (q)$ spectrum. Moreover, failure to eliminate the 
local trends leads to a wrong conclusion about the existence of
long-range correlations in this DNA walk, as evidenced by the 
overestimate  of $H$ that results from the application of  MF-DFA$0$. 
The local trends in this sequence seem to  have a particularly simple linear
nature since application of MF-DFA$1$ is sufficient to eliminate them
altogether. 

To conclude this section we note that we have considered several
alternative  one-dimensional representations, such as the strong-weak bond
classification where C or G are associated to one step up ($y_i = 1$) 
and A or T to one step down ($y_i = -1$). All the alternative schemes
analysed have produced within the numerical precision exactly the same exponents
of the purine-pyrimidine classification.
Our results could also be obtained within the  WTMM framework and, in fact, 
the wavelet analysis of the HUMHBB sequence can be found in ref. \cite{Arneodo95}. 
The aim of this section was
to give additional support to the claim of Kantelhardt {\it et al.} that,
despite the operational and conceptual simplicity of the MF-DFA method, it 
yields results that are identical to those derived by the sophisticated 
WTMM method \cite{Kantelhardt02}. For instance, this easiness of implementation has allowed us
to study  higher-dimensional DNA walks, an awkward task to be carried out using
the WTMM method, for several representations proposed
in the literature (see, e.g. \cite{Luo98,Berthelsen92,Glazier95,Oiwa02}).
As we will show in the sequel, the 
results were essentially the same as those reported here for the one-dimensional
DNA walks.

\section{DNA trails}\label{sec:3}

\begin{figure}
\centerline{\epsfig{width=0.49\textwidth,file=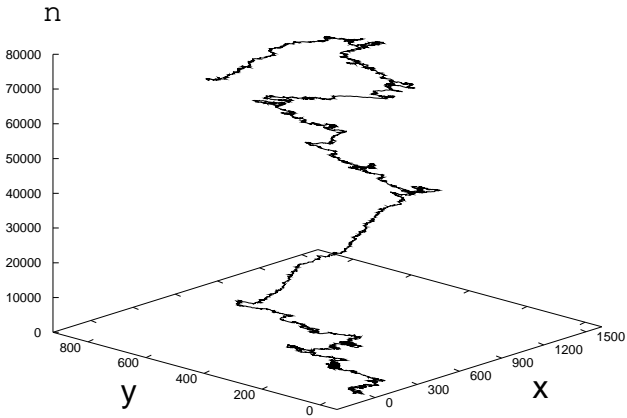}}
\par
\caption{Two-dimensional DNA-walk representation of the HUMHBB sequence whose projection
into the space plane yields the trail of Fig. \ref{fig:2}.}
\label{fig:5}
\end{figure}

As pointed out before, a DNA trail  is the projection of
a two-dimensional DNA walk into the space plane. Fig. \ref{fig:5} shows
the time record of the walk that gives rise to the trail illustrated
in Fig. \ref{fig:2} for the HUMHBB sequence. We begin by noting that the generalization
of the MF-DFA method to two or higher-dimensional records is straightforward as one
needs only to change  the basic Eq. (\ref{variancia}), which is
rewritten as 
\begin{equation}\label{var_gen}
{\hat{F}}^2 \left ( l,\nu \right ) = \frac{1}{l} \sum_{n=1}^l \left \{ {\mathbf s} \left [ \left
( \nu - 1 \right ) l + n \right ] - {\mathbf s}_{\nu} (n) \right \}^2
\end{equation}
where  ${\mathbf s} (n) = x(n) {\mathbf i} + y(n) {\mathbf j}$ and similarly for the
fitting polynomial vector ${\mathbf s}_{\nu}$.
Over an interval of $l$ steps the position vector in the trail will vary by typically 
$\mid \delta {\mathbf r} \mid \approx F_2 (l) = l^H$ while the `mass' $M$ of the trail, i.e.,
the number of points generated by these steps is $M \sim l$, provided the contribution
of overlapping points is negligible. Hence
$M \sim \mid \delta {\mathbf r} \mid^{1/H}$, from where one concludes that 
$D_t = 1/H$. This relation is valid only if $1/H < d$ where $d$ is the space dimension. 
When $1/H > d$
overlap cannot be neglected and so $D_t = d$ (see \cite{Mandelbrot82,Voss89} 
for mathematical and pictorial details of this argument). Thus knowledge of the
Hurst exponent $H$ for the $d$-dimensional DNA walks  determines uniquely
the fractal dimension of the DNA trail. Calculation of $H$ or, more generally, of
$h(q)$ follows the same procedure sketched before for the one-dimensional case, except 
that the variance given in Eq. (\ref{variancia}) is now replaced by ${\hat{F}}^2$.

Application of the MF-DFA method to two-dimensional DNA walks representing the
HUMHBB (see Fig. \ref{fig:5}) and the ECO110K  sequences yields, within
the numerical precision, the same results as for the one-dimensional case.   
Actually, so far as the value of the Hurst exponent $H$ is concerned this agreement
is expected since using Eqs. (\ref{var_gen}) and (\ref{F_q}) for $q=2$ 
we have simply $F_2 \sim l^{H_x} + l^{H_y}$, where $H_x $ and $H_y$ are the 
scaling exponents of the one-dimensional DNA walks in the plane $(n,x)$ and
$(n,y)$. Hence for large $l$ one has  $H = \mbox{max} (H_x,H_y)$.
In Table \ref{tab:dim_frac} we present the estimate of the  fractal
dimension $D_t $ of the HUMHBB and ECO110K trails, as well as of
a random trail characterized by the same length and base frequency as the 
ECO110K sequence. The errors in the estimates of $D_t$ are statistical 
errors. For the box-counting method the systematic errors
are probably much larger than those shown in  Table \ref{tab:dim_frac}, due mainly to
finite-length effects \cite{Berthelsen94}. 

Interestingly, although the $D_t$ estimates
of the MF-DFA$m$ method are based on the analysis of finite-length records, this 
has practically no effect on the results, as evidenced by the correct calculation
of $D_t$ for the random sequence, a test the box-counting method clearly fails.
Moreover, these data corroborate the intuition, stemming from the visual inspection of
Fig. \ref{fig:2}, that the effect of the local trends is 
to decrease the fractal dimension of the trail. The excellent agreement between 
the box-counting 
estimate, which is affected by both local trends and finite-length effects, 
and the MF-DFA$0$ estimate, which is affected by local trends only,
indicates that the surface effect is negligible for the HUMHBB sequence. In fact, 
this is expected since the box-counting method
is very reliable to calculate the fractal dimensions of structures that do
indeed exhibit fractality, predictably failing  in the case of space-filling structures.
The results for the
ECO110K sequence illustrate the bad performance  of box-counting when the two
sources of systematic errors are present.
The aforementioned agreement between the box-counting and the MF-DFA$0$ methods
demonstrates inequivocally that the spurious effects of local trends 
must be filtered out for a correct account of the
fractal or multifractal properties of DNA trails, a procedure that
has been ignored in all previous analyses presented in the literature 
\cite{Luo98,Gates86,Berthelsen92,Berthelsen94,Glazier95,Oiwa02}.

\begin{center}
\begin{table}
\begin{tabular} {|c|c|c|c|c|}
\colrule 
Sequence  & N      &  MF-DFA$5$       & MF-DFA$0$        & box-counting      \\ \colrule 
HUMHBB    &  73326 & $1.67 \pm 0.06$  & $1.39 \pm 0.04$  & $1.40 \pm 0.01$   \\ 
ECO110K   & 111401 & $1.96 \pm 0.08$  & $1.67 \pm 0.06$  & $1.38 \pm 0.01$   \\ 
random    & 111401 & $2.00 \pm 0.08$  & $2.00 \pm 0.08$  & $1.45 \pm 0.01$   \\  \colrule 
\end{tabular}
\caption{Estimates of the fractal dimension $D_t$ of two-dimensional trails. 
The MF-DFA$m$ estimates are based on the formula $D_t = 1/H$ where $H$ is the Hurst exponent 
of the record (e.g. Fig. \ref{fig:5}). The box-counting method is applied directly to 
the trail (e.g. Fig. \ref{fig:2}).} 
\label{tab:dim_frac}
\end{table}
\end{center}

Since the MF-DFA analysis has indicated that even the higher-dimensional DNA walks 
are monofractals, one concludes that the variance ${\hat{F}}^2 \left ( l,\nu \right )$
defined in Eq. (\ref{var_gen}) is the same for all segments $\nu$ of
the record.  As a result, the variation of the  position vector $\mid \delta {\mathbf r} \mid$
in the trail will independ of the segment $\nu$ and so the trail must have
a monofractal structure too. This conclusion seems at odds with those 
of multifractal analyses based on the sandbox and box-counting
methods \cite{Berthelsen94,Glazier95,Oiwa02}. However, we note that the original claim  
of ref. \cite{Berthelsen94} was that finite-length DNA trails, as well as randomly 
generated trails, 
are characterized by  {\it effective} multifractal spectra. Clearly, these spectra
are  artifacts of the sandbox and box-counting methods which cannot deal 
adequately with the surface effects. The unjustified dropping of
the adjective `effective' \cite{Glazier95,Oiwa02}, 
led then to the apparent disagreement between  the two approaches.
These effective multifractal spectra might indeed be an useful tool 
for comparing random or biological sequences of similar lengths \cite{Glazier95}, 
especially if one manages to free them from the local trends effects.

\section{Conclusion} \label{sec:4}

The main purpose of this contribution was to point out that the 
statistical properties
of two apparently distinct and pictorially appealing representations of DNA sequences,
namely,  DNA walks (Figs. \ref{fig:1} and \ref{fig:5})
and DNA trails (Fig. \ref{fig:2}), are in fact closely related.
In particular, we argued that the patchiness of the DNA sequences, which has
greatly hindered the characterization of the long-range correlations of DNA walks
\cite{Peng94,Karlin93},
also plagues the fractal and multifractal analyses of the DNA trails, a fact
that has been largely overlooked in previous investigations 
\cite{Berthelsen94,Glazier95,Oiwa02}. Using the recently proposed multifractal 
detrended fluctuation analysis (MF-DFA) method \cite{Kantelhardt02} 
to filter the local trends of DNA walks in $d= 1$ and $2$ dimensions, we 
were able to show that these walks are monofractals. This conclusion
holds true for the non-detrended walks as well, since the multifractal spectrum
$\tau(q)$ is linear regardless of the order $m \geq 0$ of the fitting polynomials used
to eliminate the local trends (Figs. \ref{fig:3} and \ref{fig:4}). More importantly,
the independence of the variance
$F^2 (l, \nu)$ [see Eqs. (\ref{variancia}) and (\ref{var_gen})]
on the segment $\nu$ of the DNA walk carries over to the DNA trail which is thus
monofractal too. In particular, the fractal dimension of the trail is $D_t = 1/H$
where $H$ is the Hurst exponent of the record \cite{Mandelbrot82,Voss89}. In the
case  $1/H > d$, as for the one-dimensional walks, one has $D_t = d$.
Although we have presented data for the HUMHBB  and the ECO110K sequences only, 
which  probably are, respectively, the most popular examples of intron-rich and intron-poor
sequences, we have verified that our main results hold true for  many other
sequences in the GenBank.

Once one has discarded the simple local variations in nucleotide composition
along DNA sequences as the cause of the long-range correlations observed in
intron-rich sequences (though these variations {\it do} cause the spurious long-range
correlations in intron-poor sequences), it is natural to ask then what are the 
sources for these correlations, which, as shown above, are also responsible  for
the nontrivial (fractal) geometry of the DNA trails. The answer is that neither the
internal structure of patches nor their order in the sequence
are relevant: it is the power-law distribution of patch lengths that determines the true
long-range correlations \cite{Galvan96}. This as well as several other findings 
concerning the statistical properties of DNA sequences  have prompted the 
proposal of minimal  evolutionary models,  based on biologically motivated 
mechanisms, to account  for those features 
(see, e.g., \cite{Li91,Oliveira99,Almirantis01}). In fact, rather than providing
measures to characterize or distinguish classes of sequences, we
think the  thrust of the research on  large-scale statistical properties
of genomes is to provide quantitative standards for modeling the 
inherently stochastic process of molecular evolution \cite{Kimura71}, 
bearing thus on fundamental issues such as the origin of life and the evolution 
of complexity \cite{Bonner88}.

\acknowledgments
This research was supported by Funda\c{c}\~ao de Amparo \`a Pesquisa do Estado de S\~ao Paulo 
(FAPESP), Proj.\ No.\ 99/09644-9.
The work of J.F.F. was supported in part by Conselho
Nacional de Desenvolvimento Cient\'{\i}fico e Tecnol\'ogico (CNPq).


\begin{thebibliography}{99}

\bibitem{Peng92} C.-K. Peng, S. V. Buldyrev, A. L. Goldberger, S. Havlin, 
F. Sciortino, M. Simons, and H. E. Stanley, Nature {\bf 356}, 168 (1992).

\bibitem{Peng94} C.-K. Peng, S. V. Buldyrev, S. Havlin, M. Simons, H. E. Stanley, and A. L.
Goldberger, Phys. Rev. E {\bf 49}, 1685 (1994).

\bibitem{Karlin93} S. Karlin and V. Brendel, Science {\bf 259}, 677 (1993).

\bibitem{Chui92} C. K. Chui, {\it An Introduction to Wavelets} (Academic Press, Boston, 1992).

\bibitem{Galvan96} P. Bernaola-Galv\'an, R. Rom\'an-Rold\'an, and J. L. Oliver,
Phys. Rev. E {\bf 53}, 5181 (1996).

\bibitem{Arneodo95} A. Arneodo, E. Bacry, P. V. Graves, and J. F. Muzy, 
Phys. Rev. Lett. {\bf 74}, 3293 (1995).

\bibitem{Tsonis96} A. A. Tsonis, P. Kumar, J. B. Elsner, and P. A. Tsonis, 
Phys. Rev. E {\bf 53}, 1828 (1996).

\bibitem{Luo98} L. Luo, W. Lee, L. Jia, F. Ji, and L. Tsai, Phys. Rev. E {\bf 58}, 861 (1998).

\bibitem{Mandelbrot82} B. B. Mandelbrot, {\it The Fractal Geometry of Nature}
(Freeman, San Francisco, 1982).

\bibitem{Voss89} R. F. Voss, Physica D {\bf 38}, 362 (1989).

\bibitem{Feder88} J. Feder, {\it Fractals} (Plenum Press, New York, 1988).

\bibitem{Halsey86} T. C. Halsey, M. H. Jensen, L. P. Kadanoff, I. Procaccia, and B. I.
Shraiman, Phys. Rev. A {\bf 33}, 1141 (1986).

\bibitem{Kantelhardt02} J. W. Kantelhardt, S. A. Zschiegner, E. Koscielny-Bunde, 
A. Bunde, S. Havlin, and H. E. Stanley, arXiv:physics/0202070.

\bibitem{Gates86} M. A. Gates, J. theor. Biol. {\bf 119}, 319 (1986).

\bibitem{Berthelsen92} C. L. Berthelsen, J. A. Glazier and M. H. Skolnick,
Phys. Rev. A {\bf 45}, 8902 (1992).

\bibitem{Berthelsen94} C. L. Berthelsen, J. A. Glazier, and S. Raghavachari,
Phys. Rev. E {\bf 49}, 1860 (1994).

\bibitem{Glazier95} J. A. Glazier, S. Raghavachari, C. L. Berthelsen and M. H. Skolnick, 
Phys. Rev. E {\bf 51}, 2665 (1995).

\bibitem{Oiwa02} N. N. Oiwa and J. A. Glazier, Physica A {\bf 311}, 221 (2002).

\bibitem{Vicsek92} T. Vicsek, {\it Fractal Growth Phenomena} (World Scientific,
Singapore, 1992).

\bibitem{Li91} W. Li, Phys. Rev. A {\bf 43}, 5240 (1991).

\bibitem{Oliveira99} P. M. C. de Oliveira, Physica A {\bf 273}, 70 (1999).

\bibitem{Almirantis01} Y. Almirantis and A. Provata, BioEssays {\bf 23}, 647 (2001).

\bibitem{Kimura71} M. Kimura, Theor. Pop. Biol. {\bf 2}, 174 (1971).

\bibitem{Bonner88} J. T. Bonner, {\it The Evolution of Complexity} (Princeton University
Press, Princeton, 1988).





\end{thebibliography}
\end{document}